\documentclass[sigconf,screen,nonacm]{acmart}

\AtBeginDocument{%
  
}

\usepackage{amsmath}
\usepackage{booktabs}
\usepackage{colortbl}
\usepackage{multirow}
\usepackage{tabularx}
\usepackage{pifont}
\usepackage{wasysym}
\usepackage{pgfplots}
\usepgfplotslibrary{groupplots}
\pgfplotsset{compat=1.18}

\renewcommand\footnotetextcopyrightpermission[1]{}

\title{BareWave: Waveform-Native Flow-Matching Text-to-Speech}

\author[Wei Fan]{Wei Fan\textsuperscript{1}}
\authornote{Work done during Wei Fan's internship at Tongyi Lab, Alibaba Group.}
\affiliation{%
  \institution{Anhui Province Key Laboratory of Digital Security}
  \city{Hefei}
  \country{China}
}
\email{range@mail.ustc.edu.cn}

\author[Chao-Hong Tan]{Chao-Hong Tan\textsuperscript{2}}
\authornote{Corresponding authors.}
\affiliation{%
  \institution{Tongyi Fun Team, Alibaba Group}
  \city{Hangzhou}
  \country{China}
}
\email{tanchaohong.ch@alibaba-inc.com}

\author[Qian Chen]{Qian Chen\textsuperscript{2}}
\authornotemark[2]
\affiliation{%
  \institution{Tongyi Fun Team, Alibaba Group}
  \city{Hangzhou}
  \country{China}
}
\email{tanqing.cq@alibaba-inc.com}

\author[Wen Wang]{Wen Wang\textsuperscript{2}}
\affiliation{%
  \institution{Tongyi Fun Team, Alibaba Group}
  \city{Hangzhou}
  \country{China}
}
\email{w.wang@alibaba-inc.com}

\author[Xiangang Li]{Xiangang Li\textsuperscript{2}}
\affiliation{%
  \institution{Tongyi Fun Team, Alibaba Group}
  \city{Hangzhou}
  \country{China}
}

\author[Kejiang Chen]{Kejiang Chen\textsuperscript{1}}
\authornotemark[2]
\affiliation{%
  \institution{Anhui Province Key Laboratory of Digital Security}
  \city{Hefei}
  \country{China}
}
\email{chenkj@ustc.edu.cn}

\author[Weiming Zhang]{Weiming Zhang\textsuperscript{1}}
\affiliation{%
  \institution{Anhui Province Key Laboratory of Digital Security}
  \city{Hefei}
  \country{China}
}
\email{zhangwm@ustc.edu.cn}

\author[Nenghai Yu]{Nenghai Yu\textsuperscript{1}}
\affiliation{%
  \institution{Anhui Province Key Laboratory of Digital Security}
  \city{Hefei}
  \country{China}
}
\email{ynh@ustc.edu.cn}

\makeatletter
\renewcommand{\@mkauthors}{%
  \begingroup
  \hsize=\textwidth
  \gdef\@currentauthors{}%
  \def\and{}%
  \def\@author##1{%
    \ifx\@currentauthors\@empty
      \gdef\@currentauthors{##1}%
    \else
      \g@addto@macro\@currentauthors{\quad ##1}%
    \fi
  }%
  \def\affiliation##1##2{}%
  \def\email##1##2{}%
  \addresses
  \global\setbox\mktitle@bx=\vbox{%
    \noindent\unvbox\mktitle@bx\par
    \vskip 0.9em
    \centering
    {\@authorfont \@currentauthors\par}%
    \vskip 0.65em
    {\@affiliationfont
      \textsuperscript{1}Anhui Province Key Laboratory of Digital Security\par
      \textsuperscript{2}Tongyi Fun Team, Alibaba Group\par}%
    \vskip 0.45em
    {\@affiliationfont
      \texttt{range@mail.ustc.edu.cn; \{chenkj,zhangwm,ynh\}@ustc.edu.cn}\par
      \texttt{\{tanchaohong.ch,tanqing.cq,w.wang\}@alibaba-inc.com}\par}%
    \vskip 1em
  }%
  \endgroup
}
\makeatother

\renewcommand{\shortauthors}{Fan et al.}

\begin{document}  
\setcopyright{none}
\acmDOI{}
\acmISBN{}
\acmYear{2026}
\copyrightyear{2026}
\acmConference[arXiv]{arXiv preprint}{2026}{}
\begin{abstract}
Removing intermediate representations and separately trained decoding stages has become an important direction in generative modeling.
In text-to-speech, however, high-quality systems are still commonly built through an intermediate acoustic representation before waveform synthesis.
In this work, we present BareWave, a fully waveform-native framework for direct text-to-wave generation in flow-matching TTS.
We consider this setting to raise three training challenges: raw-waveform modeling lacks a strong pretrained representational scaffold, different stages of training benefit from different noise schedules, and data-space perceptual objectives do not automatically share the temporal structure of the velocity-space flow objective.
As a result, direct waveform training is hard to optimize efficiently, hard to push toward a strong final operating point with a fixed recipe, and hard to integrate effective perceptual refinement.
Guided by this view, we develop a direct text-to-wave training framework that combines training-time representation alignment, staged noise scheduling, and velocity-aware perceptual alignment (VAPA), while preserving a single waveform-native inference path without pretrained components at test time.
Experiments on zero-shot voice cloning show that strong intelligibility, speaker similarity, and naturalness can be achieved under a fully waveform-native inference path, supporting waveform-native flow-matching TTS as a practical direction.
Project page with audio demos is available at \url{https://barewave.github.io/}.
\end{abstract}

\ccsdesc[500]{Computing methodologies~Speech synthesis}
\ccsdesc[300]{Computing methodologies~Machine learning approaches}

\keywords{text-to-speech, flow matching, waveform generation}

\maketitle
\pagestyle{plain}
\section{Introduction}
\label{sec:intro}

Removing intermediate representations and separately trained decoding stages has become an important direction in generative modeling.
Recent image generation work increasingly returns to raw pixel space and avoids tokenizers, autoencoders, or other separately trained latent interfaces, seeking more direct and unified generation pipelines~\cite{li2024arigvq,tschannen2024jetformer,yu2025pixeldit,ma2025deco,chen2025dip,wang2025pixnerd}.
In speech synthesis, the corresponding ambition is to generate waveform directly from text and prompt audio, instead of first predicting an intermediate acoustic representation and then invoking a separately trained waveform decoder.
This waveform-native direction is conceptually attractive because it promises a cleaner generation stack and a simpler deployment path.

\begin{figure}[t]
  \centering
  \includegraphics[width=0.95\linewidth]{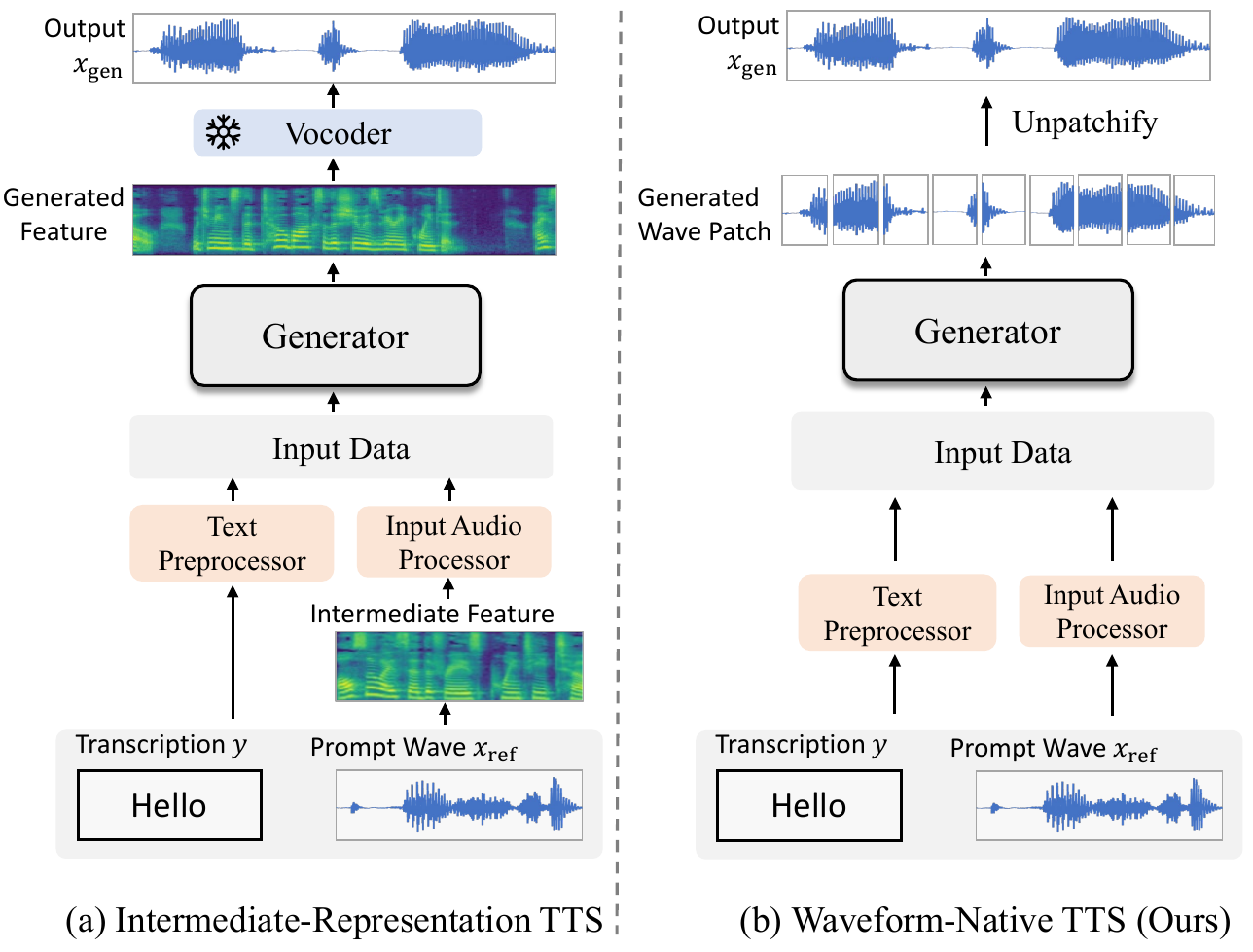}
  \caption{Mainstream TTS pipelines rely on intermediate representations and separate waveform decoders, whereas our framework is fully waveform-native.}
  \label{fig:intro_pipeline}
\end{figure}

As shown in Figure~\ref{fig:intro_pipeline}, state-of-the-art zero-shot TTS systems usually decompose synthesis into two stages: a generative model first predicts an intermediate speech representation, and a vocoder or waveform decoder renders the final waveform.
This design has become dominant because it separates high-level speech generation from low-level waveform synthesis, making quality, controllability, and zero-shot generalization easier to achieve.
Representative strong systems such as VALL-E, NaturalSpeech~3, and CosyVoice follow this pattern in different intermediate spaces~\cite{wang2023valle,ju2024naturalspeech3,du2024cosyvoice}.
Even recent simplified flow-matching systems such as E2-TTS and F5-TTS still synthesize in mel space rather than waveform space~\cite{eskimez2024e2tts,chen-etal-2025-f5}.

Direct text-to-waveform TTS pursues a stricter goal. It removes the intermediate acoustic interface itself and asks a single model to absorb both text-conditioned speech generation and waveform rendering.
Prior work has shown that this direction is feasible, and different systems have already removed different pieces of the conventional stack~\cite{weiss2021wavetacotron,gao2023e3tts,benita2023diffar}.
However, as shown in Table~\ref{tab:intro_compare}, current direct systems still usually retain some inference-time dependency, such as auxiliary acoustic or linguistic inputs, a speaker encoder, a pretrained text encoder, or a separate vocoder.
Direct TTS therefore has not yet reached a recipe that is simultaneously simple, strongly competitive, zero-shot capable, and fully waveform-native at inference.

\providecommand{\cmark}{\ding{51}}
\providecommand{\xmark}{\ding{55}}
\providecommand{\depyes}{\CIRCLE}
\providecommand{\depno}{\Circle}
\begin{table*}[t]
  \centering
  \caption{Dependency, capability and availability comparison of representative text-to-speech systems. Our framework uniquely achieves zero-shot voice cloning without inference dependencies.}
  \renewcommand{\arraystretch}{1.08}
  \setlength{\tabcolsep}{2.5pt}
  \resizebox{.9\textwidth}{!}{
  \begin{tabular}{lcccccccc}
    \toprule
    \multirow{2}{*}{Method} & \multicolumn{6}{c}{Inference-Time Dependencies} & \multicolumn{1}{c}{Reported Capability} & \multicolumn{1}{c}{Availability} \\
    \cmidrule(lr){2-7} \cmidrule(l){8-8} \cmidrule(l){9-9}
    & \shortstack[c]{Phoneme\\Inputs} & \shortstack[c]{Duration\\Predictor} & \shortstack[c]{Prosodic\\Inputs} & \shortstack[c]{Speaker\\Embedding} & \shortstack[c]{Pretrained Text\\Encoder} & \shortstack[c]{Separate\\Vocoder} & \shortstack[c]{Voice\\Cloning} & \shortstack[c]{Open\\Source} \\
    \midrule
    Wave-Tacotron~\cite{weiss2021wavetacotron} & \depno & \depno & \depno & \depno & \depno & \depno & \xmark & \xmark \\
    FastSpeech 2s~\cite{ren2021fastspeech2} & \depyes & \depyes & \depno & \depno & \depno & \depno & \xmark & \xmark \\
    DiffAR~\cite{benita2023diffar} & \depyes & \depyes & \depyes & \depno & \depno & \depno & \xmark & \cmark \\
    E3-TTS~\cite{gao2023e3tts} & \depno & \depno & \depno & \depyes & \depyes & \depno & \cmark & \xmark \\
    E2-TTS~\cite{eskimez2024e2tts} & \depno & \depno & \depno & \depno & \depno & \depyes & \cmark & \xmark \\
    F5-TTS~\cite{chen-etal-2025-f5} & \depno & \depno & \depno & \depno & \depno & \depyes & \cmark & \cmark \\
    Ours & \depno & \depno & \depno & \depno & \depno & \depno & \cmark & \cmark* \\
    \bottomrule
  \end{tabular}}
  \par
  \raggedright\footnotesize \hspace{3.8em} * Our code and checkpoints will be released soon.
  \label{tab:intro_compare}
\end{table*}

For waveform-native TTS, a more specific question is: \emph{once these inference-time components are removed, what training design is needed to make the resulting system work well?}
We consider this setting to raise three training challenges.
First, raw-waveform modeling lacks a strong pretrained representational scaffold, so the model must organize linguistic, speaker, and acoustic structure in a space with much weaker priors.
Second, different stages of training benefit from different noise schedules: early optimization favors a convergence-friendly distribution, whereas later refinement benefits from greater coverage of cleaner regions.
Third, waveform-native generation requires data-space perceptual refinement for spectral quality, yet the velocity-space flow objective gives the main loss a time-dependent structure that auxiliary perceptual objectives do not automatically share.
These factors make direct waveform TTS hard to optimize efficiently under matched budgets, hard to push toward a strong final operating point, and hard to integrate effective perceptual refinement with the velocity-space flow objective.

Guided by this view, we present BareWave, a system that keeps the inference path minimal and places all additional complexity in training design.
We use training-time representation alignment to provide the missing speech prior. Specifically, we align the hidden states of the generator with features extracted from a frozen self-supervised speech model. To better match the needs of optimization as training evolves, we combine staged noise scheduling with velocity-aware perceptual alignment. The early training phase utilizes a convergence-friendly logit-normal noise-level distribution. In the subsequent stage, we switch to a uniform distribution that assigns higher sampling density to cleaner regions. Concurrently, we introduce Velocity-Aware Perceptual Alignment (VAPA), which scales a multi-resolution spectral distance so that perceptual refinement reflects the temporal structure of the velocity-space flow objective.
All auxiliary branches are removed at test time, so the model remains a single waveform-native generator.

Experiments on zero-shot voice cloning show that strong speech synthesis can be achieved under a fully waveform-native inference path, without intermediate acoustic representations, pretrained inference-time components, or a separate vocoder.
Ablation results further indicate that performance in this setting is strongly shaped by training design, including persistent training-time representation guidance, staged noise scheduling, and velocity-aware perceptual alignment.
Taken together, these results show that waveform-native flow-matching TTS can support strong zero-shot voice cloning under a single direct text-to-wave inference path.

Our main contributions are as follows:
\begin{itemize}
	\item We present BareWave, a fully waveform-native framework for zero-shot TTS, with no intermediate acoustic representation, pretrained inference-time component, or separate vocoder.
	\item We develop a direct text-to-wave training recipe that combines representation alignment, staged noise scheduling, and velocity-aware perceptual alignment while preserving a single waveform-native inference path.
  \item We observe that data-space perceptual losses are implicitly down-weighted relative to the velocity-space flow objective near cleaner timesteps, and propose Velocity-Aware Perceptual Alignment (VAPA) to match the temporal structure of the flow objective.                                           
                                                                  
\end{itemize}

\section{Related Work}
\label{sec:related}

\subsection{Waveform-Space Speech Generation}

Waveform generation for speech is commonly studied in two related settings: vocoding and direct text-to-waveform synthesis.
In the vocoder setting, a model is given an intermediate acoustic representation such as a mel-spectrogram and is asked to render waveform samples.
In the direct TTS setting, the model must instead absorb both speech generation and waveform rendering into a single system.

Neural vocoders are important because they isolate waveform modeling from linguistic planning, and this setting has established many effective waveform-level architectures and objectives.
Representative examples include adversarial vocoders such as Parallel WaveGAN and HiFi-GAN, as well as diffusion-style waveform generators such as DiffWave and WaveGrad~\cite{yamamoto2020parallelwavegan,kong2020hifigan,kong2021diffwave,chen2021wavegrad}.
However, a vocoder is not itself a full text-to-speech system, because it assumes that an upstream model has already produced the intermediate acoustic representation.

Direct text-to-waveform TTS is therefore a more demanding problem, because the model must jointly solve text-conditioned speech generation and waveform synthesis.
Prior work such as Wave-Tacotron, DiffAR, and E3-TTS has shown that direct waveform synthesis is a viable TTS route, while also revealing that strong performance often still relies on additional structural support or conditioning signals~\cite{weiss2021wavetacotron,benita2023diffar,gao2023e3tts}, leaving a fully waveform-native zero-shot recipe unresolved.

\subsection{Modern Zero-Shot TTS Pipelines}

Zero-shot TTS studies whether a model can synthesize speech for a previously unseen speaker from text and a short reference utterance, without speaker-specific finetuning.
Recent progress has made this setting much stronger and more practical through large-scale generative modeling.
Representative systems include codec-based and latent-space pipelines such as VALL-E, NaturalSpeech~3, and CosyVoice, as well as simplified flow-matching systems such as E2-TTS and F5-TTS~\cite{wang2023valle,ju2024naturalspeech3,du2024cosyvoice,eskimez2024e2tts,chen-etal-2025-f5}.
This literature shows that explicit aligners, duration predictors, and heavily engineered text frontends are no longer always necessary for high-quality zero-shot synthesis.
However, the dominant recipe in this literature therefore still relies on intermediate representations, typically together with a pretrained codec model or a separate waveform decoder at inference, so whether comparable zero-shot performance can be achieved under a fully waveform-native inference path remains less explored.

\subsection{Representation Alignment for Raw-Space Generation}

Representation alignment uses external pretrained features as training targets or priors, and is especially appealing when a generator must learn in a space whose structure is hard to recover directly.
In image generation, REPA showed that aligning noisy hidden states with pretrained representations can improve the optimization of diffusion and flow models in raw space~\cite{yu2025repa}.
In speech, self-supervised models such as HuBERT and WavLM have become widely used across downstream tasks, making them natural sources of pretrained structure~\cite{hsu2021hubert,chen2022wavlm}.
Accordingly, recent TTS work has begun to introduce such models into training-time alignment, for example through dual-modality alignment for accelerating diffusion-based synthesis~\cite{choi2025adma}.
For direct waveform TTS, where the intermediate acoustic interface is removed, representation alignment is a natural way to inject pretrained speech structure during training; however, how such alignment should be used as a training-only support in direct waveform TTS remains underexplored.

\section{Method}
\label{sec:method}

\subsection{Overview}

\begin{figure*}[t]
  \centering
  \includegraphics[width=\textwidth]{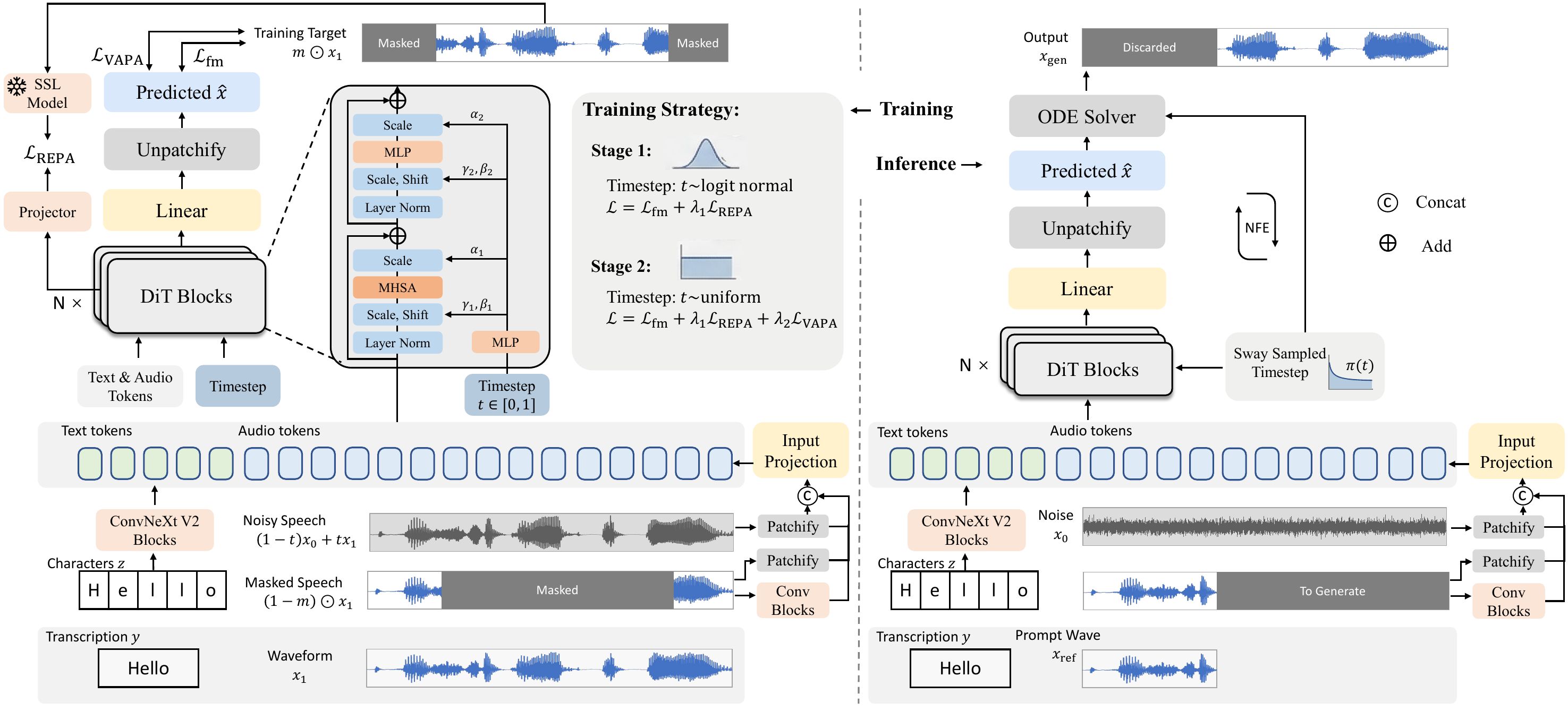}
  \caption{Overview of the proposed waveform-native TTS framework. In the training stage, the generator is guided by a main flow-matching branch and an auxiliary representation alignment branch. The optimization utilizes a staged schedule with distinct timestep ($t$) distributions, where velocity-aware perceptual alignment is enabled only after the model reaches the refinement stage. At inference, all auxiliary components are discarded, preserving a single direct text-to-waveform generator.}
  \Description{Diagram placeholder for the waveform-native generator, training-only alignment branch, and staged training schedule.}
  \label{fig:framework}
\end{figure*}

We study zero-shot direct text-to-waveform synthesis with a waveform-native generator.
Figure~\ref{fig:framework} gives an overview of the proposed waveform-native TTS framework.
Given input text $c$ and a short prompt waveform $a$, the model directly produces the target waveform $\hat{x}$ without an intermediate acoustic representation or a separate vocoder at inference.
Our architecture is a waveform-patch DiT backbone trained with conditional flow matching.
The inference path is kept unchanged, while the extra support is introduced only during training through representation alignment, staged noise scheduling, and velocity-aware perceptual alignment.

\subsection{Preliminaries}
\label{subsec:preliminaries}

\subsubsection{Flow Matching}

We formulate training as a conditional speech-infilling problem~\cite{chen-etal-2025-f5}.
A random target span is selected as the region to generate, while the remaining waveform serves as the masked-speech context together with the full text condition.
Consider a clean distribution $x \sim p_{\mathrm{data}}(x)$ and a noise distribution $\epsilon \sim p_{\mathrm{noise}}(\epsilon)$.
During training, the noisy state $z_t$ is defined by the standard linear interpolation path~\cite{lipman2023flowmatching,li2025backtobasics}:
\begin{equation}
z_t = t x + (1-t)\epsilon, \qquad t \in [0,1],
\label{eq:corruption}
\end{equation}
where $t$ is the continuous time variable, so that $z_t$ starts from the noise endpoint at $t=0$ and reaches the data endpoint at $t=1$.
The corresponding flow velocity is the time derivative of $z_t$,
\begin{equation}
v = \frac{\partial z_t}{\partial t} = x-\epsilon.
\label{eq:velocity}
\end{equation}
Conditional flow matching learns a velocity field $v_\theta(z_t,t,c,a)$ under the text condition $c$ and the prompt-audio condition $a$ by minimizing:
\begin{equation}
\mathcal{L}_{\mathrm{CFM}}
=
\mathbb{E}_{x,\epsilon,t}
\left[
\left\|
v_\theta(z_t,t,c,a)-v
\right\|_2^2
\right].
\label{eq:cfm}
\end{equation}
At inference time, generation is performed by integrating the sampling ODE from $t=0$ to $t=1$.

\subsubsection{Prediction Space}

The prediction space specifies which quantity the network outputs directly~\cite{li2025backtobasics}.
In our model, the network is parameterized to directly predict the clean waveform target:
\begin{equation}
\hat{x}_\theta = f_\theta(z_t, t, c, a).
\label{eq:xpred}
\end{equation}
That is, the prediction target is the clean waveform $x$ rather than the noise $\epsilon$ or the velocity $v$.

\subsubsection{Loss Space}

While the loss is typically defined in a reference space, it can equivalently be expressed in other spaces.
With a fixed reparameterization from one prediction space to another, this change induces an effective reweighting of the objective~\cite{li2025backtobasics}.
For the $x$-prediction and $v$-loss combination, we have:
\begin{equation}
\hat{v}_\theta = \frac{\hat{x}_\theta-z_t}{1-t},
\qquad
v = \frac{x-z_t}{1-t},
\label{eq:xpred_vloss}
\end{equation}
where $\hat{v}_\theta$ is the induced velocity prediction converted from the network output $\hat{x}_\theta$, and $v$ is the target velocity induced by the clean waveform $x$ and the noisy state $z_t$.
For speech infilling, the masked objective is written as:
\begin{equation}
\mathcal{L}_{\mathrm{fm}}
=
\mathbb{E}_{i:\,M_i=1}
\bigl(\hat{v}_{\theta,i}-v_i\bigr)^2
=
\mathbb{E}_{i:\,M_i=1}
\frac{1}{(1-t)^2}
\bigl(\hat{x}_{\theta,i}-x_i\bigr)^2,
\label{eq:fm_loss}
\end{equation}
where $\mathbb{E}_{i:\,M_i=1}$ denotes the average over waveform samples selected by the binary infilling mask $M$, making Eq.~\eqref{eq:fm_loss} a reweighted form of the corresponding $x$-loss.
In implementation, whenever an $x$-prediction is converted to the velocity-space quantity in Eq.~\eqref{eq:xpred_vloss}, we clip the denominator from below, using $\bar{d}_t=\max(1-t,\epsilon_t)$, to avoid numerical instability near the clean endpoint.

\subsubsection{Sampling and Classifier-Free Guidance}

At sampling time, the waveform trajectory is generated by integrating the induced velocity field.
The sampling ODE is:
\begin{equation}
\frac{\mathrm{d} z_t}{\mathrm{d} t} = \tilde{v}_\theta(z_t, t, c, a).
\label{eq:ode}
\end{equation}
In practice, we solve this ODE with a fixed-step Heun solver.
Classifier-free guidance is enabled by randomly dropping text, prompt-audio, or both conditions during training to create conditional and unconditional branches.
Given the conditional velocity $\hat{v}_t^{\mathrm{cond}}$ and the unconditional velocity $\hat{v}_t^{\mathrm{uncond}}$, inference uses:
\begin{equation}
\tilde{v}_t
=
\hat{v}_t^{\mathrm{uncond}}
+
s(t)\bigl(\hat{v}_t^{\mathrm{cond}}-\hat{v}_t^{\mathrm{uncond}}\bigr),
\label{eq:cfg}
\end{equation}
where $s(t)$ equals the guidance strength within a prescribed guidance interval and reduces to $1$ outside that interval.
This suggests that the guidance only modifies the velocity field inside a specified time interval and otherwise falls back to the conditional branch.

\subsection{Waveform-Patch DiT Backbone}
\label{subsec:backbone}

Our generator operates on waveform patches rather than spectrogram frames or codec tokens.
The noisy waveform $z_t$ is first patchified by a 1-D convolutional embedder whose kernel size and stride both match the waveform patch size.
This produces a sequence of waveform patch tokens that serves as the main signal stream of the transformer.
To condition on the prompt audio $a$, a dedicated convolutional frontend extracts coarse, downsampled acoustic features aligned to the target patch rate, which are then concatenated with a parallel stream of direct raw-waveform prompt patches.
The text condition is represented at the character level, refined by lightweight ConvNeXt-style blocks, and inserted as in-context tokens before the audio stream.
This yields a single token sequence in which text provides left-context guidance, while the noisy waveform and the hierarchical prompt-audio representations remain in the main generation stream.
The resulting sequence is processed by DiT blocks with timestep modulation and rotary positional encoding.
The final layer predicts patch-wise clean waveform values, and a non-overlapping unpatchify operation maps the patch predictions back to the sample domain.
The entire architecture is trained end-to-end from scratch; pre-trained models are utilized strictly to provide auxiliary supervision during training rather than being loaded as architectural components.
As a result, the inference path remains a single waveform-native generator from text and prompt audio to waveform samples, and the remaining design choices, described below, provide optimization support without introducing additional inference-time components.
\vspace{-1mm}

\subsection{Training Strategy}
\label{subsec:training_strategy}

\subsubsection{Training-Time Representation Alignment}
\label{subsec:input_align}

To provide a speech prior during training, we align a selected hidden state of the generator to frozen self-supervised speech features.
Concretely, we extract a hidden representation $h_\ell$ from transformer layer $\ell$ and compare it against a frozen WavLM\footnote{https://huggingface.co/microsoft/wavlm-base-plus} teacher sequence $h^{\mathrm{teacher}}$ computed from the ground-truth target waveform.
Before the comparison, $h_\ell$ is linearly interpolated to the teacher length and passed through a lightweight alignment head $\phi(\cdot)$.
Following prior dual-modality alignment designs~\cite{choi2025adma}, this head consists of two Conv1d--GroupNorm--Mish blocks followed by a $1\times1$ Conv1d output layer.
The alignment loss is defined as:
\begin{equation}
\mathcal{L}_{\mathrm{REPA}}
=
\mathbb{E}_{i \in \Omega}
\left[
1 - \cos\!\left(\phi(h_\ell)_i, h^{\mathrm{teacher}}_i\right)
\right],
\label{eq:align}
\end{equation}
where $\mathbb{E}_{i \in \Omega}$ denotes the average over valid aligned positions shared by the projected source sequence and the teacher sequence.
This branch acts only as a training-time prior and does not alter the generator architecture itself.
At inference, the teacher network and the projection head are removed.

\subsubsection{Staged Noise Scheduling}
\label{subsec:t_schedule}

Direct waveform TTS is highly sensitive to the noise-level distribution used during flow-matching training.
Different sampling distributions over noise levels emphasize different denoising subproblems and lead to different operating points.
We therefore do not rely on a single fixed noise schedule throughout optimization.
Instead, we use a two-stage schedule indexed by normalized training progress $u \in [0,1]$.
The first stage uses a convergence-friendly logit-normal noise-level distribution, and the second stage switches to a broader uniform continuation once the model has reached a workable waveform solution, which better supports learning to denoise from cleaner noise states, further improving detail quality.
\begin{equation}
p(t \mid u) =
\begin{cases}
p_{\mathrm{logit\text{-}normal}}(t;\mu,\sigma), & u < \rho, \\
p_{\mathrm{uniform}}(t), & u \ge \rho,
\end{cases}
\label{eq:t_schedule}
\end{equation}
where $\rho$ denotes the stage-switch point.
In practice, the first stage is parameterized by the logit-normal sampling family together with its $(P_{\mathrm{mean}}, P_{\mathrm{std}})$ setting, while the continuation stage switches the noise-level distribution to uniform.

\subsubsection{Velocity-Aware Perceptual Alignment}
\label{subsec:spectral}

Waveform-native generation benefits from perceptual losses that explicitly constrain spectral structure.
In the $x$-prediction, $v$-loss framework adopted here, the flow-matching objective is evaluated in velocity space while the perceptual objective is defined in data space.
Figure~\ref{fig:velocity_scaling_schematic} illustrates the corresponding temporal scaling from an $x$-space quantity to its velocity-space form.
\definecolor{scaleblue}{RGB}{55,118,186}
\definecolor{scalegray}{RGB}{92,92,92}
\definecolor{noisyshade}{RGB}{240,244,250}
\definecolor{cleanshade}{RGB}{231,239,252}

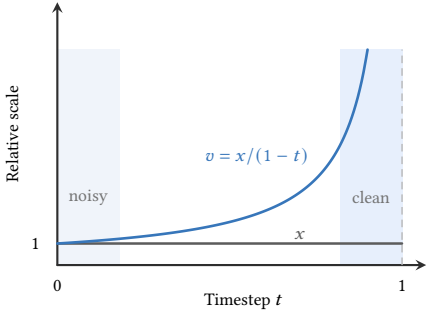
\begin{figure}[t]
  \centering
\scalebox{0.95}{
\begin{tikzpicture}[x=4.8cm,y=0.30cm,>=stealth]
  \fill[noisyshade] (0,0) rectangle (0.18,10.0);
  \fill[cleanshade] (0.82,0) rectangle (1,10.0);
  \node[font=\footnotesize, text=black!55, anchor=center] at (0.09,3.15) {noisy};
  \node[font=\footnotesize, text=black!55, anchor=center] at (0.91,3.15) {clean};

  \draw[->, line width=0.85pt, black!82, line cap=round] (0,0) -- (1.07,0);
  \draw[->, line width=0.85pt, black!82, line cap=round] (0,0) -- (0,12.2);

  \draw[densely dashed, line width=0.55pt, black!28] (1,0) -- (1,10.0);
  \node[font=\footnotesize, anchor=north] at (0,-0.30) {$0$};
  \node[font=\footnotesize, anchor=north] at (1,-0.30) {$1$};
  \node[font=\footnotesize, anchor=east] at (-0.028,1) {$1$};
  \node[font=\footnotesize, anchor=north] at (0.54,-0.95) {Timestep $t$};
  \node[font=\footnotesize, rotate=90, anchor=south] at (-0.09,6.1) {Relative scale};

  \draw[scalegray, line width=1.0pt, line cap=round] (0,1) -- (1,1);
  \node[font=\footnotesize, text=scalegray, anchor=south west, inner sep=1pt] at (0.68,1.08) {$x$};

  \draw[
    scaleblue,
    line width=1.05pt,
    line cap=round,
    smooth,
  ] plot coordinates {
(0.0000,1.0000) (0.0114,1.0115) (0.0228,1.0233) (0.0342,1.0354) (0.0456,1.0477) (0.0570,1.0604) (0.0684,1.0734) (0.0797,1.0867) (0.0911,1.1003) (0.1025,1.1142) (0.1139,1.1286) (0.1253,1.1433) (0.1367,1.1584) (0.1481,1.1738) (0.1595,1.1898) (0.1709,1.2061) (0.1823,1.2229) (0.1937,1.2402) (0.2051,1.2580) (0.2165,1.2763) (0.2278,1.2951) (0.2392,1.3145) (0.2506,1.3345) (0.2620,1.3551) (0.2734,1.3763) (0.2848,1.3982) (0.2962,1.4209) (0.3076,1.4442) (0.3190,1.4684) (0.3304,1.4934) (0.3418,1.5192) (0.3532,1.5460) (0.3646,1.5737) (0.3759,1.6024) (0.3873,1.6322) (0.3987,1.6632) (0.4101,1.6953) (0.4215,1.7287) (0.4329,1.7634) (0.4443,1.7995) (0.4557,1.8372) (0.4671,1.8765) (0.4785,1.9175) (0.4899,1.9603) (0.5013,2.0051) (0.5127,2.0519) (0.5241,2.1011) (0.5354,2.1526) (0.5468,2.2067) (0.5582,2.2636) (0.5696,2.3235) (0.5810,2.3867) (0.5924,2.4534) (0.6038,2.5240) (0.6152,2.5987) (0.6266,2.6780) (0.6380,2.7622) (0.6494,2.8520) (0.6608,2.9478) (0.6722,3.0502) (0.6835,3.1600) (0.6949,3.2780) (0.7063,3.4052) (0.7177,3.5426) (0.7291,3.6916) (0.7405,3.8537) (0.7519,4.0306) (0.7633,4.2246) (0.7747,4.4382) (0.7861,4.6746) (0.7975,4.9375) (0.8089,5.2318) (0.8203,5.5634) (0.8316,5.9398) (0.8430,6.3710) (0.8544,6.8696) (0.8658,7.4528) (0.8772,8.1443) (0.8886,8.9773) (0.9000,10.0000)
  };
  \node[font=\footnotesize, text=scaleblue, anchor=west, inner sep=1pt] at (0.42,5.05) {$v=x/(1-t)$};
\end{tikzpicture}
}
  \caption{Velocity-space scaling relative to a data-space quantity. The horizontal reference denotes \(x\), and the curve denotes \(v=x/(1-t)\), which grows as \(t\to1\).}
  \Description{A schematic plot with timestep t on the horizontal axis and relative scale on the vertical axis. The left and right sides are lightly shaded as noisy and clean, respectively. The data-space quantity x stays at scale 1, while v equals x divided by 1 minus t and increases toward t equals 1.}
  \label{fig:velocity_scaling_schematic}
\end{figure}
As Eq.~\eqref{eq:fm_loss} shows, the velocity-space evaluation applies an implicit $1/(1-t)^2$ factor to data-space prediction errors, increasing their effective weight near cleaner timesteps.
With a fixed nominal coefficient, a data-space perceptual loss does not carry this factor, so its relative weight against the main flow objective decreases near the clean endpoint, weakening its role in the low-noise refinement region emphasized by the flow loss.

We address this with Velocity-Aware Perceptual Alignment (VAPA).
Following WaveFM~\cite{luo-etal-2025-wavefm}, we use a multi-resolution STFT perceptual distance $\mathcal{D}_{\mathrm{STFT}}(\hat{x},x)$ that combines phase, log-magnitude, local spectral-gradient, and Laplacian terms across multiple resolutions over the generated region.
VAPA applies a velocity-aware temporal scaling to this perceptual distance:
\begin{equation}
\mathcal{L}_{\mathrm{VAPA}}
=
\mathbb{E}_{x,\epsilon,t}
\left[
(1-t)^{-\gamma}
\mathcal{D}_{\mathrm{STFT}}\bigl(\hat{x}_\theta(z_t,t,c,a),x\bigr)
\right].
\label{eq:vapa}
\end{equation}
Here $\gamma$ controls the scaling strength. We use $\gamma=1$, following the norm order of the base L1 STFT and mel distances; the same scaling is applied uniformly across all resolution and refinement terms. With $\gamma=1$, the perceptual weight grows as $1/(1-t)$ near the clean endpoint, corresponding to the temporal scaling that an L1 $x$-space distance acquires under the $x$-to-$v$ loss-space conversion.
For numerical stability, the implementation follows the same denominator-clipping convention as the flow-matching MSE conversion above: $(1-t)$ is replaced by $\bar{d}_t=\max(1-t,\epsilon_t)$.
This term is enabled only in the refinement stage ($u \ge \rho$).

\subsection{Training and Inference Objectives}
\label{subsec:train_infer}

The complete objective combines the basic flow-matching loss with the optional representation-alignment and velocity-aware perceptual terms:
\begin{equation}
\mathcal{L}_{\mathrm{total}}
=
\mathcal{L}_{\mathrm{fm}}
+ \lambda_1 \mathcal{L}_{\mathrm{REPA}}
+ \mathbb{I}[u \ge \rho] \, \lambda_2 \mathcal{L}_{\mathrm{VAPA}}.
\label{eq:total}
\end{equation}
Here $\lambda_1$ and $\lambda_2$ control the strengths of the representation-alignment and velocity-aware perceptual terms, respectively, and $\mathbb{I}[u \ge \rho]$ denotes the late-stage switch.
The generator itself is shared across all stages, and only the training objective changes over the course of optimization.
At inference, all training-only branches are removed, and the model keeps only the text in-context path, the concatenative prompt-audio path, and the waveform-patch generator.
This yields a clean waveform-native inference graph with no intermediate acoustic representation and no pretrained component in the test-time path.

\section{Experiments}
\label{sec:experiments}

\newcommand{\expbreaksubsubsection}[1]{\subsubsection{#1}\mbox{}\par}
\newcommand{\expinlinehead}[1]{\noindent\textbf{#1} }

\begin{table*}[t]
  \centering
  \renewcommand{\arraystretch}{1.06}
  \caption{Zero-shot voice cloning results on Seed-TTS test-en and LibriSpeech(-PC) test-clean, measured by WER (\%), SIM-o, and UTMOS. Training data size and inference-time parameter count are shown for comparability.}
  \label{tab:main_results}
  \resizebox{.9\linewidth}{!}{
  \begin{tabular}{lccccc ccc}
    \toprule
    \multirow{2}{*}{Method} & \multirow{2}{*}{Training data} & \multirow{2}{*}{Params} &
    \multicolumn{3}{c}{Seed-TTS test-en} & \multicolumn{3}{c}{LibriSpeech-PC} \\
    \cmidrule(lr){4-6} \cmidrule(lr){7-9}
    & & & WER (\%) $\downarrow$ & SIM-o $\uparrow$ & UTMOS $\uparrow$ & WER (\%) $\downarrow$ & SIM-o $\uparrow$ & UTMOS $\uparrow$ \\
    \midrule
    \textbf{Ground Truth} & -- & -- & 1.86 & 0.734 & 3.53 & 2.48 & 0.695 & 4.10 \\
    \midrule
    \rowcolor[RGB]{242,244,255}
    \multicolumn{9}{l}{\textit{Larger-data intermediate-representation references}} \\
    \hspace{1em}CosyVoice 2~\cite{du2024cosyvoice2} & 167K Multi. & 618M & 2.51 & 0.659 & 4.15 & 2.05 & 0.659 & 4.38 \\
    \hspace{1em}FireRedTTS~\cite{chen-etal-2025-f5}$^\dagger$ & 248K Multi. & $\sim$580M & 3.82 & 0.46 & -- & 2.69 & 0.47 & -- \\
    \midrule
    \rowcolor[RGB]{242,244,255}
    \multicolumn{9}{l}{\textit{Same-data intermediate-representation baselines}} \\
    \hspace{1em}F5-TTS Base~\cite{chen-etal-2025-f5} (+ Vocos) & 19.4k EN & 335.8M + 13.5M & \underline{2.09} & 0.573 & \textbf{3.83} & \underline{3.17} & 0.597 & \textbf{4.10} \\
    \hspace{1em}E2-TTS Base~\cite{eskimez2024e2tts} (+ Vocos) & 19.4k EN & 333M + 13.5M & 3.50 & \underline{0.582} & 3.41 & 4.32 & \textbf{0.632} & 3.84 \\
    \rowcolor[RGB]{242,244,255}
    \multicolumn{9}{l}{\textit{Same-data waveform-native systems}} \\
    \hspace{1em}Simple direct-wave baseline & 19.4k EN & 983.6M & 2.42 & 0.424 & 3.35 & 3.45 & 0.416 & 3.60 \\
    \hspace{1em}Ours (basic training) & 19.4k EN & 983.6M & 2.34 & 0.478 & 3.43 & 3.32 & 0.471 & 3.69 \\
    \hspace{1em}Ours (proposed training scheme) & 19.4k EN & 983.6M & \textbf{1.75} & \textbf{0.602} & \underline{3.72} & \textbf{2.88} & \underline{0.614} & \underline{4.01} \\
    \bottomrule
  \end{tabular}
  }
  \par
  \vspace{0.2ex}
  \begin{minipage}{\linewidth}
    \raggedright\footnotesize
    \hspace{3.5em}$^\dagger$ Numbers are taken from prior reported results. Bold and underline mark the best and second-best results among same-data systems.
  \end{minipage}
\end{table*}

\vspace{-0.3em}
\subsection{Experimental Setup}

\expbreaksubsubsection{Training Data and Task}
In our experiments, the controlled same-data systems are trained on the English subset of the in-the-wild speech dataset Emilia~\cite{he2024emilia}. We filter out utterances with transcription failures or misclassified language labels, retain 19.4k hours of audio at 24kHz, and the evaluation task is zero-shot voice cloning from text and a short prompt utterance. Larger-data public systems are included as external benchmarks and are reported with their original training-data scale in Table~\ref{tab:main_results}.

\expbreaksubsubsection{Evaluation Protocol and Metrics}

For English objective evaluation, we follow the same protocol used by F5-TTS~\cite{chen-etal-2025-f5}.
We report WER (\%) for intelligibility and content fidelity, SIM-o for speaker identity preservation, and UTMOS for perceptual naturalness.
WER is computed with Whisper-large-v3 implemented through Faster-Whisper and reported as a percentage, SIM-o is computed with a WavLM-large-based ECAPA-TDNN speaker verification model, and UTMOS is obtained from the open-source SpeechMOS evaluator.

\expbreaksubsubsection{Baselines}

\expinlinehead{Intermediate-representation baselines.}
For same-data intermediate-representation baselines, we include F5-TTS~\cite{chen-etal-2025-f5} and E2-TTS~\cite{eskimez2024e2tts}.
We train their Base configurations on the same 19.4k-hour Emilia English subset used by our proposed method, and evaluate them with the default Vocos vocoder~\cite{siuzdak2023vocos} and default inference settings.
For larger-data reference systems, we include CosyVoice 2~\cite{du2024cosyvoice2}, evaluated by running inference from its public checkpoint, and FireRedTTS values transcribed from the F5-TTS paper~\cite{chen-etal-2025-f5}.
Reported-source numbers are marked with a dagger in Table~\ref{tab:main_results}.

\expinlinehead{Waveform-native baselines.}
We include two direct-wave baselines to separate prompt-encoder gains from training-recipe gains.
For the simple direct-wave baseline, we condition directly on raw waveform patches without the convolutional prompt frontend and train the model only with $\mathcal{L}_{\mathrm{fm}}$ under a fixed logit-normal schedule.
For the basic-training baseline, we add the dual-branch prompt encoder described in Section~\ref{subsec:backbone} but keep the
same minimal training recipe.
Our full model retains this encoder and adds the proposed training design: representation alignment, late
uniform continuation, and velocity-aware perceptual alignment.

\expbreaksubsubsection{Implementation Details}

\expinlinehead{Model architecture and optimization.}
Our waveform-native backbone uses 32 DiT blocks with 16 attention heads and hidden size 1280, resulting in approximately 983.6M total parameters.
During training, we use Muon with Adam~\cite{jordan2024muon} on 4 NVIDIA A800 GPUs, with Muon learning rate $10^{-3}$ for DiT parameters and Adam learning rate $5\times10^{-5}$ with betas $(0.9, 0.95)$ for the remaining parameters.
Weight decay is set to zero, and the learning rate remains constant after 20k warmup updates.
For the models without REPA loss, we use batch size of 19200 audio frames, and for the models with REPA loss, we use batch size of 9600 audio frames due to memory constraints. To account for this difference, all update counts reported hereafter are expressed as equivalent updates, normalized to match the total data volume seen by the base model without REPA.

\expinlinehead{Training schedule.}
Our first training stage uses a logit-normal distribution over $t$, where $\mathrm{logit}(t)\sim\mathcal{N}(\mu,\sigma^2)$.
We sample $s\sim\mathcal{N}(\mu,\sigma^2)$ and set $t=\mathrm{sigmoid}(s)$.
All waveform-native runs use noise scale 0.1, with $\mu=-0.4$ and $\sigma=0.8$ in this first stage, and then switch to uniform continuation in the later stage.
We train all the models for 640k equivalent updates. For REPA runs with uniform continuation, we switch to uniform at 240k updates.
Following~\cite{chen-etal-2025-f5}, we first drop the masked speech input with a rate of 0.3, then drop the masked speech again but with text input together with a rate of 0.2~\cite{le2023voicebox} during training.

\expinlinehead{Training losses.}
The REPA branch uses $\lambda_1=0.0025$.
When velocity-aware perceptual alignment is enabled, the multi-resolution STFT distance uses $\lambda_2=4\times10^{-4}$ with the WaveFM-style refined STFT setting, FFT sizes $[1024, 2048, 512]$, hop sizes $[128,256,64]$, window sizes $[512,1024,256]$, $\gamma=1$, and $\epsilon_t=0.01$.
The refined distance follows WaveFM~\cite{luo-etal-2025-wavefm} in combining phase, log-magnitude, local time/frequency spectral-gradient, and Laplacian terms. It is the base perceptual distance used in the mechanism ablation; VAPA differs by applying the velocity-aware scaling above.

\expinlinehead{EMA and Inference settings.}
We maintain two EMA tracks with decay rates 0.9999 and 0.9996 throughout training, and we use the 0.9999 track for inference.
During inference, we use 50 NFE steps and the full sampling interval $[0,1]$.
We also apply the sway sampling technique with coefficient $s=-1.0$~\cite{chen-etal-2025-f5}, and a classifier-free guidance scale of 3.5.

\begin{figure*}[t]
  \centering
  \begin{tikzpicture}
\begin{groupplot}[
group style={group size=3 by 1, horizontal sep=0.78cm},
width=0.31\textwidth,
height=0.23\textwidth,
xlabel={Equivalent Steps (k)},
xtick={80,160,240,320,400},
grid=major,
grid style={line width=.1pt, draw=gray!25},
major grid style={line width=.2pt,draw=gray!35},
tick label style={font=\small},
label style={font=\small},
title style={font=\small},
legend style={font=\footnotesize, draw=none, fill=none, inner xsep=1pt, inner ysep=1pt, row sep=0pt},
]
\nextgroupplot[
title={WER (\%)},
ylabel={Metric value},
legend columns=1,
legend pos=north east,
legend style={font=\scriptsize, draw=none, fill=white, fill opacity=0.85, text opacity=1, inner sep=1.5pt, row sep=0pt}
]
\addplot+[mark=*, line width=0.9pt, mark size=1.8pt, color={rgb,255:red,31; green,119; blue,180}] coordinates {(80,21.62) (160,7.66) (240,5.92) (320,4.66) (400,4.12)};
\addlegendentry{w/o REPA}
\addplot+[mark=square*, line width=0.9pt, mark size=1.8pt, color={rgb,255:red,214; green,39; blue,40}] coordinates {(80,9.44) (160,5.03) (240,3.84) (320,3.29) (400,3.19)};
\addlegendentry{REPA}
\nextgroupplot[title={SIM-o}]
\addplot+[mark=*, line width=0.9pt, mark size=1.8pt, color={rgb,255:red,31; green,119; blue,180}] coordinates {(80,0.259) (160,0.354) (240,0.393) (320,0.420) (400,0.437)};
\addplot+[mark=square*, line width=0.9pt, mark size=1.8pt, color={rgb,255:red,214; green,39; blue,40}] coordinates {(80,0.385) (160,0.452) (240,0.486) (320,0.502) (400,0.513)};
\nextgroupplot[title={UTMOS}]
\addplot+[mark=*, line width=0.9pt, mark size=1.8pt, color={rgb,255:red,31; green,119; blue,180}] coordinates {(80,2.88) (160,3.33) (240,3.45) (320,3.54) (400,3.60)};
\addplot+[mark=square*, line width=0.9pt, mark size=1.8pt, color={rgb,255:red,214; green,39; blue,40}] coordinates {(80,3.40) (160,3.53) (240,3.65) (320,3.71) (400,3.70)};
\end{groupplot}
\end{tikzpicture}
  \caption{Comparison of training curves with and without REPA on LibriSpeech(-PC) test-clean. Both models are trained on the same Emilia English subset. Using REPA provides a strong speech prior, which improves data efficiency and speaker similarity.}
  \Description{Three small line charts showing WER (\%), SIM-o, and UTMOS for REPA and non-REPA base models versus aligned seen-data budget.}
  \label{fig:repa_curves}
\end{figure*}
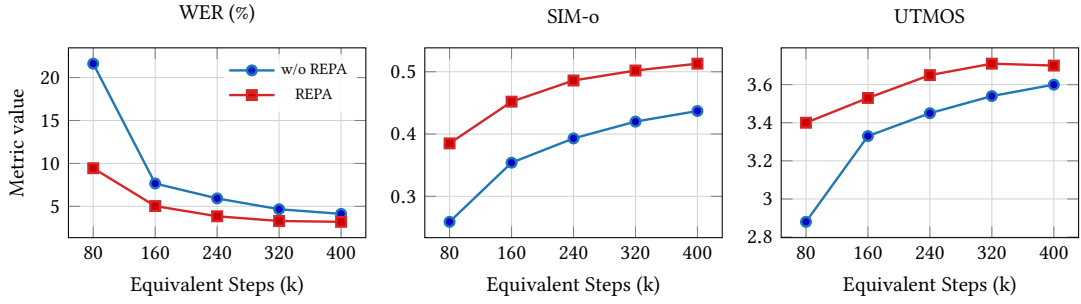

\begin{figure*}[t]
  \centering
  \definecolor{curveblue}{RGB}{31,119,180}
\definecolor{curvered}{RGB}{214,39,40}
\begin{tabular}[b]{@{}c@{\hspace{0.90cm}}c@{}}
\begin{tikzpicture}[baseline=(current bounding box.south)]
\begin{axis}[
scale only axis,
width=3.45cm,
height=2.75cm,
grid=major,
grid style={line width=.1pt, draw=gray!25},
major grid style={line width=.2pt,draw=gray!35},
tick label style={font=\small},
label style={font=\small},
title style={font=\small},
title={Density $p(t)$},
xlabel={$t$},
ylabel={Density},
xlabel style={yshift=-1pt},
ylabel style={at={(axis description cs:-0.18,0.5)}},
xmin=0,
xmax=1,
ymin=0,
ymax=2.8,
xtick={0,0.25,0.5,0.75,1},
]
\addplot[
  draw=none,
  fill=curveblue,
  fill opacity=0.28,
] coordinates {
(0.0010,0.0000) (0.0093,0.0005) (0.0176,0.0087) (0.0259,0.0386) (0.0343,0.0980) (0.0426,0.1873) (0.0509,0.3025) (0.0592,0.4376) (0.0675,0.5866) (0.0759,0.7438) (0.0842,0.9043) (0.0925,1.0642) (0.1008,1.2204) (0.1091,1.3705) (0.1174,1.5128) (0.1258,1.6461) (0.1341,1.7695) (0.1424,1.8825) (0.1507,1.9849) (0.1590,2.0768) (0.1673,2.1582) (0.1757,2.2294) (0.1840,2.2907) (0.1923,2.3426) (0.2006,2.3855) (0.2089,2.4199) (0.2172,2.4461) (0.2256,2.4649) (0.2339,2.4765) (0.2422,2.4815) (0.2505,2.4804) (0.2588,2.4736) (0.2671,2.4616) (0.2754,2.4447) (0.2838,2.4234) (0.2921,2.3981) (0.3004,2.3690) (0.3087,2.3366) (0.3170,2.3012) (0.3253,2.2631) (0.3337,2.2225) (0.3420,2.1797) (0.3503,2.1350) (0.3586,2.0886) (0.3669,2.0407) (0.3752,1.9916) (0.3836,1.9415) (0.3919,1.8904) (0.4002,1.8387) (0.4085,1.7863) (0.4168,1.7336) (0.4252,1.6806) (0.4335,1.6275) (0.4418,1.5743) (0.4501,1.5212) (0.4584,1.4683) (0.4667,1.4157) (0.4751,1.3635) (0.4834,1.3117) (0.4917,1.2605) (0.5000,1.2099) (0.5083,1.1599) (0.5166,1.1107) (0.5250,1.0622) (0.5333,1.0146) (0.5416,0.9679) (0.5499,0.9220) (0.5582,0.8772) (0.5665,0.8334) (0.5748,0.7906) (0.5832,0.7488) (0.5915,0.7082) (0.5998,0.6686) (0.6081,0.6302) (0.6164,0.5929) (0.6247,0.5569) (0.6331,0.5220) (0.6414,0.4882) (0.6497,0.4557) (0.6580,0.4244) (0.6663,0.3944) (0.6746,0.3655) (0.6830,0.3378) (0.6913,0.3114) (0.6996,0.2862) (0.7079,0.2622) (0.7162,0.2394) (0.7245,0.2179) (0.7329,0.1975) (0.7412,0.1782) (0.7495,0.1602) (0.7578,0.1433) (0.7661,0.1275) (0.7744,0.1128) (0.7828,0.0992) (0.7911,0.0867) (0.7994,0.0752) (0.8077,0.0648) (0.8160,0.0553) (0.8244,0.0467) (0.8327,0.0391) (0.8410,0.0323) (0.8493,0.0263) (0.8576,0.0211) (0.8659,0.0167) (0.8743,0.0129) (0.8826,0.0098) (0.8909,0.0072) (0.8992,0.0051) (0.9075,0.0035) (0.9158,0.0023) (0.9242,0.0014) (0.9325,0.0008) (0.9408,0.0004) (0.9491,0.0002) (0.9574,0.0001) (0.9657,0.0000) (0.9741,0.0000) (0.9824,0.0000) (0.9907,0.0000) (0.9990,0.0000)
} \closedcycle;
\addplot[
  draw=none,
  fill=curvered,
  fill opacity=0.18,
  domain=0:1,
  samples=2
] {1} \closedcycle;
\node[
  anchor=north east,
  fill=white,
  fill opacity=0.90,
  text opacity=1,
  inner sep=2.0pt,
  rounded corners=1pt
] at (rel axis cs:0.97,0.97) {
  \begin{tikzpicture}[x=1cm,y=1cm,baseline=(current bounding box.north)]
    \fill[curveblue, opacity=0.28] (0.00,0.34) rectangle (0.15,0.48);
    \node[anchor=west, font=\footnotesize] at (0.20,0.41) {Logit-normal};
    \fill[curvered, opacity=0.18] (0.00,0.02) rectangle (0.15,0.16);
    \node[anchor=west, font=\footnotesize] at (0.20,0.09) {Uniform};
  \end{tikzpicture}
};
\end{axis}
\end{tikzpicture}
&
\begin{tikzpicture}[baseline=(current bounding box.south)]
\begin{groupplot}[
group style={group size=3 by 1, horizontal sep=0.55cm},
scale only axis,
width=3.25cm,
height=2.75cm,
xlabel={Equivalent Steps (k)},
grid=major,
grid style={line width=.1pt, draw=gray!25},
major grid style={line width=.2pt,draw=gray!35},
tick label style={font=\small},
label style={font=\small},
title style={font=\small},
legend style={font=\scriptsize, draw=none, fill=white, fill opacity=0.85, text opacity=1, inner sep=1.5pt, row sep=0pt},
enlarge x limits=0.08,
enlarge y limits=0.10,
]
\nextgroupplot[
title={WER (\%)},
ylabel={Metric value},
ylabel style={at={(axis description cs:-0.14,0.5)}},
xlabel style={yshift=-1pt},
xmin=75,
xmax=405,
ymin=0,
ymax=102,
xtick={80,160,240,320,400},
legend columns=1,
legend pos=north east
]
\addplot+[mark=*, line width=0.9pt, mark size=1.8pt, color=curveblue]
coordinates {(80,21.62) (160,7.66) (240,5.92) (320,4.66) (400,4.12)};
\addlegendentry{Logit-normal}
\addplot+[mark=square*, line width=0.9pt, mark size=1.8pt, color=curvered]
coordinates {(80,97.17) (160,17.32) (240,8.43) (320,5.80) (400,4.28)};
\addlegendentry{Uniform}

\nextgroupplot[
title={SIM-o},
xlabel style={yshift=-1pt},
xmin=75,
xmax=405,
ymin=0.05,
ymax=0.56,
xtick={80,160,240,320,400},
]
\addplot+[mark=*, line width=0.9pt, mark size=1.8pt, color=curveblue]
  coordinates {(80,0.259) (160,0.354) (240,0.393) (320,0.420) (400,0.437)};
\addplot+[mark=square*, line width=0.9pt, mark size=1.8pt, color=curvered]
  coordinates {(80,0.105) (160,0.271) (240,0.321) (320,0.350) (400,0.365)};

\nextgroupplot[
title={UTMOS},
xlabel style={yshift=-1pt},
xmin=75,
xmax=405,
ymin=2.0,
ymax=3.85,
xtick={80,160,240,320,400},
]
\addplot+[mark=*, line width=0.9pt, mark size=1.8pt, color=curveblue]
  coordinates {(80,2.88) (160,3.33) (240,3.45) (320,3.54) (400,3.60)};
\addplot+[mark=square*, line width=0.9pt, mark size=1.8pt, color=curvered]
  coordinates {(80,2.17) (160,3.24) (240,3.49) (320,3.68) (400,3.74)};
\end{groupplot}
\end{tikzpicture}
\end{tabular}
  \caption{Analysis of early noise-level distributions evaluated on LibriSpeech(-PC) test-clean. Both models are trained on the same Emilia English subset without REPA from scratch. The left panel shows the density of the timestep $t$ for both logit-normal and uniform distributions. The remaining panels show the training curves for WER (\%), SIM-o, and UTMOS. The logit-normal schedule focuses on middle noise levels and leads to faster early optimization than the uniform distribution.}
  \Description{A four-panel figure in a single row. The first panel shows shaded densities for the adopted logit-normal noise-level distribution and a uniform distribution over t, with a small in-panel legend. The remaining three panels show WER (\%), SIM-o, and UTMOS versus training updates for the same two noise-level distributions. A line legend appears in the upper right of the WER panel.}
  \label{fig:corruption_regimes}
\end{figure*}
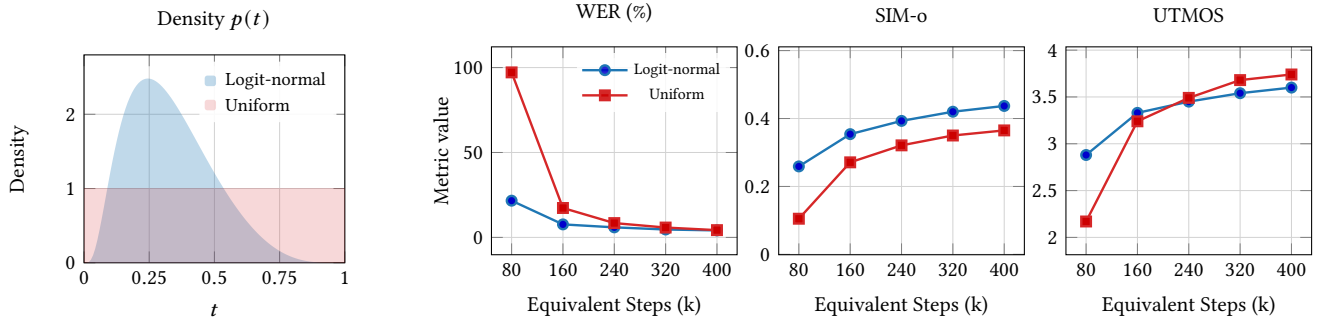

\subsection{Main Results}

Table~\ref{tab:main_results} reports zero-shot voice-cloning results on Seed-TTS test-en and LibriSpeech-PC. Project page with audio demos is available at \url{https://barewave.github.io/}.

\subsubsection{Comparison with intermediate-representation systems.} Among same-data systems, BareWave achieves competitive results. On Seed-TTS test-en, BareWave reaches 1.75\% WER, 0.602 SIM-o, and 3.72 UTMOS, giving the best WER and SIM-o among same-data systems. On LibriSpeech-PC, BareWave reaches 2.88\% WER and 0.614 SIM-o, improving over F5-TTS Base on both metrics, though trailing E2-TTS Base on SIM-o.
These results suggest that a waveform-native inference path achieves strong content intelligibility, and that speaker-related structure can be organized effectively in waveform space.
BareWave remains slightly below F5-TTS Base on UTMOS, which we attribute to the inherent difficulty of single-stage waveform rendering without a dedicated vocoder.

\subsubsection{Comparison with waveform-native systems.} We next examine the same-data waveform-native systems.
The simplest configuration predicts waveform patches directly from raw waveform patch prompts.

We find that the simple direct-wave baseline already reaches 3.45\% WER, 0.416 SIM-o, and 3.60 UTMOS.
This indicates that direct waveform modeling is already capable of producing intelligible speech, and that the synthesized speech begins to capture the target speaker characteristics.
A second baseline introduces a stronger prompt-wave frontend and improves to 3.32\% WER, 0.471 SIM-o, and 3.69 UTMOS, showing that a stronger prompt-side representation helps waveform-native TTS reach a better operating point.
Compared with our basic-training waveform-native baseline, the full training recipe substantially narrows the remaining gap to F5-TTS.
Specifically, representation alignment provides an additional training-time representational scaffold, while staged noise scheduling and velocity-aware perceptual alignment drive further improvements in speaker similarity and naturalness.
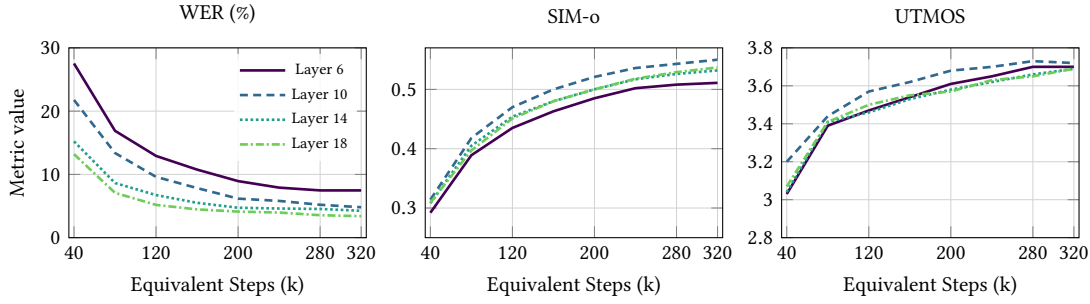
\begin{figure*}[t]
  \centering
  \definecolor{repaLsix}{RGB}{68,1,84}
\definecolor{repaLten}{RGB}{49,104,142}
\definecolor{repaLfourteen}{RGB}{31,158,137}
\definecolor{repaLeighteen}{RGB}{109,205,89}

\begin{tikzpicture}
\begin{groupplot}[
group style={group size=3 by 1, horizontal sep=0.78cm},
width=0.31\textwidth,
height=0.23\textwidth,
xlabel={Equivalent Steps (k)},
xmin=35,
xmax=325,
xtick={40,120,200,280,320},
grid=major,
grid style={line width=.1pt, draw=gray!25},
major grid style={line width=.2pt,draw=gray!35},
tick label style={font=\small},
label style={font=\small},
title style={font=\small},
legend style={font=\scriptsize, draw=none, fill=none, inner xsep=1pt, inner ysep=1pt, row sep=0pt},
legend columns=4,
]

\nextgroupplot[
title={WER (\%)},
ylabel={Metric value},
ymin=0,
ymax=30,
legend columns=1,
legend pos=north east,
legend style={font=\scriptsize, draw=none, fill=white, fill opacity=0.85, text opacity=1, inner sep=1.5pt, row sep=0pt}
]
\addplot+[mark=none, line width=1.0pt, color=repaLsix] coordinates {(40,27.53) (80,16.89) (120,12.92) (160,10.76) (200,8.94) (240,7.91) (280,7.47) (320,7.47)};
\addlegendentry{Layer 6}
\addplot+[mark=none, line width=1.0pt, densely dashed, color=repaLten] coordinates {(40,21.77) (80,13.38) (120,9.61) (160,7.84) (200,6.17) (240,5.81) (280,5.20) (320,4.81)};
\addlegendentry{Layer 10}
\addplot+[mark=none, line width=1.0pt, densely dotted, color=repaLfourteen] coordinates {(40,15.22) (80,8.64) (120,6.73) (160,5.51) (200,4.73) (240,4.61) (280,4.54) (320,4.25)};
\addlegendentry{Layer 14}
\addplot+[mark=none, line width=1.0pt, dash pattern=on 3pt off 1.1pt on 0.8pt off 1.1pt, color=repaLeighteen] coordinates {(40,13.20) (80,7.09) (120,5.18) (160,4.47) (200,4.12) (240,3.97) (280,3.55) (320,3.40)};
\addlegendentry{Layer 18}

\nextgroupplot[
title={SIM-o},
ymin=0.25,
ymax=0.57
]
\addplot+[mark=none, line width=1.0pt, color=repaLsix] coordinates {(40,0.292) (80,0.389) (120,0.435) (160,0.463) (200,0.485) (240,0.502) (280,0.508) (320,0.511)};
\addplot+[mark=none, line width=1.0pt, densely dashed, color=repaLten] coordinates {(40,0.314) (80,0.418) (120,0.470) (160,0.500) (200,0.521) (240,0.536) (280,0.543) (320,0.550)};
\addplot+[mark=none, line width=1.0pt, densely dotted, color=repaLfourteen] coordinates {(40,0.308) (80,0.405) (120,0.454) (160,0.480) (200,0.500) (240,0.517) (280,0.526) (320,0.532)};
\addplot+[mark=none, line width=1.0pt, dash pattern=on 3pt off 1.1pt on 0.8pt off 1.1pt, color=repaLeighteen] coordinates {(40,0.309) (80,0.397) (120,0.451) (160,0.480) (200,0.500) (240,0.518) (280,0.529) (320,0.537)};

\nextgroupplot[
title={UTMOS},
ymin=2.8,
ymax=3.8
]
\addplot+[mark=none, line width=1.0pt, color=repaLsix] coordinates {(40,3.03) (80,3.39) (120,3.47) (160,3.54) (200,3.61) (240,3.65) (280,3.70) (320,3.70)};
\addplot+[mark=none, line width=1.0pt, densely dashed, color=repaLten] coordinates {(40,3.20) (80,3.44) (120,3.57) (160,3.62) (200,3.68) (240,3.70) (280,3.73) (320,3.72)};
\addplot+[mark=none, line width=1.0pt, densely dotted, color=repaLfourteen] coordinates {(40,3.04) (80,3.41) (120,3.46) (160,3.53) (200,3.58) (240,3.62) (280,3.66) (320,3.69)};
\addplot+[mark=none, line width=1.0pt, dash pattern=on 3pt off 1.1pt on 0.8pt off 1.1pt, color=repaLeighteen] coordinates {(40,3.07) (80,3.41) (120,3.50) (160,3.55) (200,3.57) (240,3.63) (280,3.65) (320,3.69)};
\end{groupplot}
\end{tikzpicture}
  \caption{Effect of the REPA alignment target layer on training dynamics. The curves trace WER, SIM-o, and UTMOS over equivalent steps for different teacher layers. Deeper teacher layers reduce WER, while the intermediate layer 10 yields stronger SIM-o than the shallow layer-6 target, showing a metric trade-off across alignment targets. All runs use the same first-stage logit-normal setting ($P_{\mathrm{mean}}=-0.4$, $P_{\mathrm{std}}=0.8$) and are evaluated on LibriSpeech-PC.}
  \Description{Three line charts showing WER, SIM-o, and UTMOS over equivalent steps for different REPA teacher layers.}
  \label{fig:repa_layer_sensitivity}
\end{figure*}

\begin{figure*}[t]
  \centering
  \input{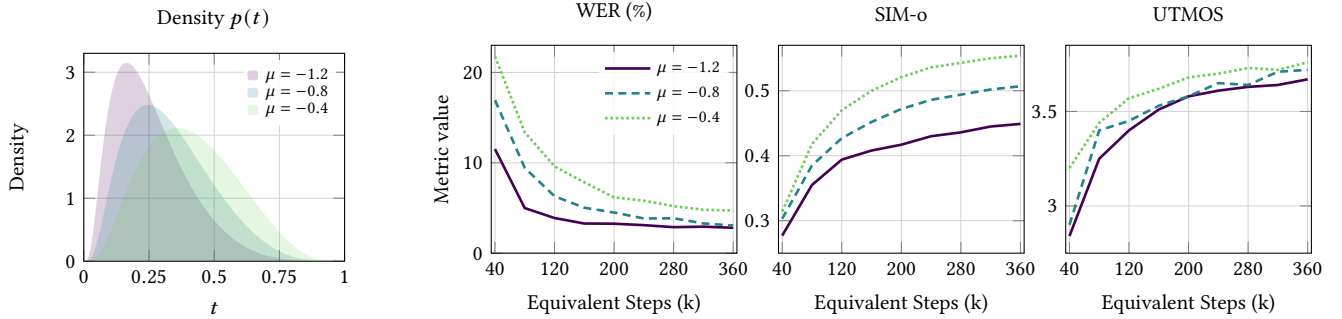}
  \caption{Effect of the logit-normal mean $\mu$ on the noise-level distribution and training dynamics. The density panel shows how $\mu$ shifts the sampled noise levels, and the remaining panels trace WER, SIM-o, and UTMOS over equivalent steps. Lower $\mu$ gives lower WER in this sweep, while higher $\mu$ improves SIM-o and UTMOS. This indicates that the early noise-level distribution changes the model's operating point rather than yielding a uniformly best setting. All runs use REPA layer 10, $P_{\mathrm{std}}=0.8$, and are evaluated on LibriSpeech-PC.}
  \Description{A four-panel figure showing shaded logit-normal density areas for different means, followed by WER, SIM-o, and UTMOS curves over equivalent steps.}
  \label{fig:logit_mean_sensitivity}
\end{figure*}

\begin{table}[t]
  \centering
  \renewcommand{\arraystretch}{1.06}
  \setlength{\tabcolsep}{4.2pt}
  \caption{Comparison results of different training recipes on LibriSpeech(-PC) test-clean near the 600k equivalent training budget.}
  \label{tab:ablation}
  \resizebox{\linewidth}{!}{
  \begin{tabular}{lccc}
    \toprule
    Recipe & WER (\%) $\downarrow$ & SIM-o $\uparrow$ & UTMOS $\uparrow$ \\
    \midrule
    Base (w/o REPA) & 3.32 & 0.471 & 3.69 \\
    + REPA & 2.86 & 0.522 & 3.70 \\
    \quad + late uniform & 2.93 & 0.543 & 3.82 \\
    \quad + late uniform + refined STFT & 3.38 & 0.585 & 3.88 \\
    \rowcolor[RGB]{242,244,255}
    \quad + late uniform + refined STFT + VAPA & 3.52 & 0.610 & 3.97 \\
    \bottomrule
  \end{tabular}}
\end{table}

\subsection{Ablation Study}

\subsubsection{REPA Ablation}

Figure~\ref{fig:repa_curves} isolates training-time representation alignment by comparing REPA and non-REPA base models under matched seen-data budgets.
The results show that REPA improves data efficiency throughout the training process. The most stable improvement is seen in SIM-o, while the model also reaches better WER and UTMOS points under the same data budget.
This behavior is consistent with the role of REPA in our method, namely to provide a stronger speech prior during training and make it easier to organize representation in a waveform-native model.

\subsubsection{Noise Schedule Ablation}

Table~\ref{tab:ablation} reports matched-budget recipe variants for the noise-schedule and perceptual-loss ablations. Building on the REPA-supported model, late uniform continuation yields 2.93\% WER, 0.543 SIM-o, and 3.82 UTMOS (versus 2.86\%, 0.522, and 3.70 without continuation). The continuation stage therefore improves speaker similarity and naturalness while keeping WER in a narrow range.

\subsubsection{Perceptual Loss Ablation}

With late uniform continuation fixed, adding the refined STFT distance improves SIM-o from 0.543 to 0.585 and UTMOS from 3.82 to 3.88. Applying VAPA to the same refined spectral distance further improves SIM-o to 0.610 and UTMOS to 3.97, while WER remains in the same operating range, indicating that velocity-aware scaling makes the spectral distance more effective in the continuation stage.

\subsection{Analysis}

\subsubsection{Early Noise-Level Distributions}

The logit-normal noise-level distribution is important for fast early convergence.
Figure~\ref{fig:corruption_regimes} compares two from-scratch models that differ only in their noise-level distributions. While the uniform distribution puts more weight on cleaner states, the logit-normal schedule focuses on middle noise levels. This difference in density leads to a clear gap in training speed. The logit-normal model converges much faster in both WER (\%) and SIM-o within the same training budget, while UTMOS is less strongly affected by the early optimization gap. These results show that the early stage benefits from the logit-normal noise-level distribution for faster optimization.

\subsubsection{Input-Side Representation Hierarchy}

Prompt-side representation quality is a key factor in waveform-native TTS. Table~\ref{tab:input_bottleneck} compares two prompt representation variants using the same training data. The simple baseline uses only waveform patches from the prompt audio. In contrast, the version with representation hierarchy adds a downsampled prompt stream to the raw prompt path, creating a representation hierarchy. The results show that this stronger hierarchy improves both WER (\%) and SIM-o. This suggests that providing more prompt-side information helps improve waveform-native performance.

\begin{table}[t]
  \centering
  \renewcommand{\arraystretch}{1.06}
  \setlength{\tabcolsep}{7.2pt}
  \caption{Evaluation of the input-side representation hierarchy on LibriSpeech(-PC) test-clean. Both models are trained on the same Emilia English subset without REPA. The results show that representation hierarchy improves both WER (\%) and SIM-o.}
  \label{tab:input_bottleneck}
  \begin{tabular}{lcc}
    \toprule
    Variant & WER (\%) $\downarrow$ & SIM-o $\uparrow$ \\
    \midrule
    Simple direct-wave baseline & 3.45 & 0.416 \\
    \rowcolor[RGB]{242,244,255}
    + Representation Hierarchy & 3.32 & 0.471 \\
    \bottomrule
  \end{tabular}
\end{table}

\subsubsection{REPA Layer Sensitivity}

Figure~\ref{fig:repa_layer_sensitivity} compares different teacher layers for training-time representation alignment across matched equivalent-step checkpoints. The sweep shows that the alignment target changes the balance between intelligibility, speaker similarity, and naturalness: deeper layers reduce WER, while layer 10 yields the strongest SIM-o. This analysis shows that the alignment target introduces a trade-off across metrics rather than uniformly improving all dimensions.

\subsubsection{Logit-Normal Mean Sensitivity}

Figure~\ref{fig:logit_mean_sensitivity} compares different means for the logit-normal noise-level distribution across matched equivalent-step checkpoints. Lower means favor WER in this sweep, while higher means improve SIM-o and UTMOS. The result shows that the early noise-level distribution changes the operating point of the model, rather than producing a single uniformly best setting across all metrics.

\subsubsection{Classifier-Free Guidance Strength.}

\definecolor{cfgbase}{RGB}{88,124,170}
\definecolor{cfgbaseedge}{RGB}{57,86,123}

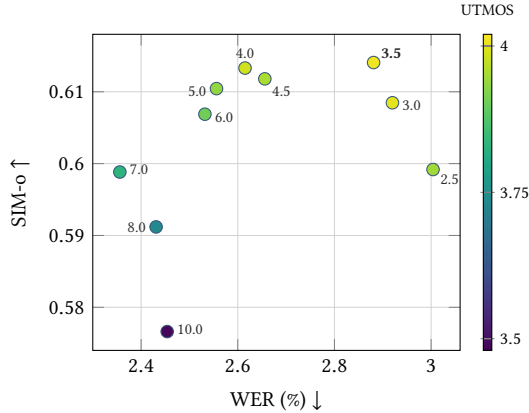
\begin{figure}[t]
  \centering
\begin{tikzpicture}
\begin{axis}[
width=0.76\columnwidth,
height=0.68\columnwidth,
xmin=2.50,
xmin=2.30,
xmax=3.06,
ymin=0.574,
ymax=0.618,
xlabel={WER (\%) $\downarrow$},
ylabel={SIM-o $\uparrow$},
xlabel style={yshift=-1pt},
ylabel style={yshift=1pt},
grid=major,
grid style={line width=.1pt, draw=gray!25},
major grid style={line width=.2pt,draw=gray!35},
tick label style={font=\small},
label style={font=\small},
title style={font=\small},
clip=false,
xtick={2.40,2.60,2.80,3.00},
ytick={0.58,0.59,0.60,0.61},
scaled x ticks=false,
xticklabel style={/pgf/number format/fixed,/pgf/number format/precision=2},
colormap/viridis,
colorbar,
point meta min=3.48,
point meta max=4.02,
colorbar style={
  width=3.2pt,
  title={UTMOS},
  title style={font=\scriptsize, yshift=-2pt},
  tick label style={font=\scriptsize},
  ytick={3.5,3.75,4.0},
  yticklabel style={/pgf/number format/fixed,/pgf/number format/precision=2},
},
]
\addplot[
  scatter,
  only marks,
  scatter src=explicit,
  mark=*,
  mark size=2.35pt,
  scatter/use mapped color={draw=cfgbaseedge, fill=mapped color},
] table[meta=utmos] {
wer sim utmos
3.004 0.59918 3.9446
2.920 0.60847 3.9998
2.881 0.61407 4.0081
2.615 0.61330 3.9819
2.656 0.61180 3.9520
2.556 0.61044 3.9327
2.532 0.60688 3.9026
2.356 0.59883 3.8290
2.431 0.59119 3.7331
2.454 0.57662 3.4887
};

\node[font=\scriptsize, text=black!82, inner sep=0.6pt, anchor=north west, xshift=2pt, yshift=-1pt]
  at (axis cs:3.004,0.59918) {2.5};
\node[font=\scriptsize, text=black!82, inner sep=0.6pt, anchor=west, xshift=3pt, yshift=-1pt]
  at (axis cs:2.920,0.60847) {3.0};
\node[font=\scriptsize\bfseries, text=black!85, inner sep=0.6pt, anchor=south west, xshift=2pt, yshift=1pt]
  at (axis cs:2.881,0.61407) {3.5};
\node[font=\scriptsize, text=black!82, inner sep=0.6pt, anchor=south, yshift=3pt]
  at (axis cs:2.615,0.61330) {4.0};
\node[font=\scriptsize, text=black!82, inner sep=0.6pt, anchor=north west, xshift=2pt, yshift=-2pt]
  at (axis cs:2.656,0.61180) {4.5};
\node[font=\scriptsize, text=black!82, inner sep=0.6pt, anchor=east, xshift=-3pt, yshift=-1pt]
  at (axis cs:2.556,0.61044) {5.0};
\node[font=\scriptsize, text=black!82, inner sep=0.6pt, anchor=west, xshift=3pt, yshift=-1pt]
  at (axis cs:2.532,0.60688) {6.0};
\node[font=\scriptsize, text=black!82, inner sep=0.6pt, anchor=west, xshift=3pt, yshift=1pt]
  at (axis cs:2.356,0.59883) {7.0};
\node[font=\scriptsize, text=black!82, inner sep=0.6pt, anchor=east, xshift=-3pt]
  at (axis cs:2.431,0.59119) {8.0};
\node[font=\scriptsize, text=black!82, inner sep=0.6pt, anchor=west, xshift=3pt, yshift=1pt]
  at (axis cs:2.454,0.57662) {10.0};

\end{axis}
\end{tikzpicture}
  \caption{Comparison of different CFG strength values under the Heun-50 bf16 inference setting. Marker color indicates UTMOS. Stronger guidance shifts the operating point toward lower WER but eventually lowers both SIM-o and UTMOS. We select CFG 3.5 as the main operating point because it gives a good balance between the metrics.}
  \label{fig:cfg_pareto_wer_sim}
\end{figure}
We further analyze the effect of CFG strength at inference. We scan over a CFG range from 2.5 to 10.0 under the Heun-50 bf16 setting and plot the WER and SIM-o results in Figure~\ref{fig:cfg_pareto_wer_sim}. Stronger CFG generally improves WER, but SIM-o peaks at CFG 3.5 before declining under stronger guidance.
The trend suggests that stronger CFG helps improve content intelligibility, while excessive guidance harms speaker feature preservation.
In the main experiments, we choose CFG 3.5 because it achieves the best SIM-o and UTMOS in this sweep while maintaining competitive WER.

\section{Limitations}
\label{sec:discussion}

Our waveform-native generator uses 983.6\,M parameters, larger than the compared mel-based baselines, leading to higher training and inference cost. Architectural compression and distillation are promising directions for future work.
\section{Conclusion}
\label{sec:conclusion}

In this paper, we presented BareWave, a fully waveform-native zero-shot voice cloning framework. To address the three training challenges of direct waveform TTS, namely the lack of representational priors, the absence of a universally optimal noise schedule, and the temporal mismatch between data-space perceptual objectives and the velocity-space flow objective, we introduced a training scheme comprising representation alignment, staged noise scheduling, and Velocity-Aware Perceptual Alignment (VAPA). All auxiliary components are removed at inference, preserving a single direct text-to-waveform path. Experiments show that this recipe achieves competitive results relative to strong mel-based baselines while maintaining a strictly waveform-native inference path. These findings support waveform-native flow-matching TTS as a practical direction for zero-shot speech synthesis.


\bibliographystyle{ACM-Reference-Format}
\bibliography{acmart}

\appendix
\newpage
\section{Detailed Experimental Setup}

\newcommand{\suppinlinehead}[1]{\par\addvspace{0.1\baselineskip}\noindent\textbf{#1} }

\subsection{Training Data and Preprocessing}

All waveform-native runs in this work are trained on the English subset of Emilia.
We build the training set after filtering out utterances with transcription failure or language mismatch.
After filtering, the resulting English corpus contains 19.4k hours of speech.

During training, we retain only samples whose durations lie between 0.3~s and 30~s.
Each utterance is converted to mono when necessary and resampled to 24~kHz.
For waveform input, we remove only the DC offset before batching and do not apply additional normalization.

\subsection{Detailed Evaluation Protocol}

\suppinlinehead{Cross-sentence evaluation protocol.}
For English objective evaluation, we use a cross-sentence zero-shot voice-cloning protocol on LibriSpeech-PC test-clean.
We use the released 4-to-10-second LibriSpeech-PC test-clean subset with 1127 evaluation samples\footnote{https://github.com/swivid/f5-tts}.
Each target utterance is paired with a same-speaker prompt utterance under the cross-sentence setting.

\suppinlinehead{Inference input construction.}
For each evaluation pair, the synthesis text is formed by concatenating the prompt transcript and the target transcript.
Unless otherwise specified, the target duration is estimated from the prompt duration and the prompt/target text-length ratio, avoiding oracle duration information at test time.

\subsection{Implementation Details}

\begin{table}[H]
  \centering
  \small
  \renewcommand{\arraystretch}{1.05}
  \newcommand{\tparam}[1]{\hspace{0.6em}#1}
  \caption{Waveform-patch DiT backbone settings used for our experiments.}
  \label{tab:supp_arch}
  \begin{tabularx}{\linewidth}{@{}p{0.48\linewidth}X@{}}
    \toprule
    Parameter & Value \\
    \midrule
    \multicolumn{2}{@{}l}{\textit{Waveform-patch DiT backbone}} \\
    \tparam{Patch size} & 768 samples \\
    \tparam{Patch stride} & 768 samples \\
    \tparam{Bottleneck dimension} & 768 \\
    \tparam{Depth} & 32 \\
    \tparam{Hidden size} & 1280 \\
    \tparam{Attention heads} & 16 \\
    \tparam{MLP ratio} & 4.0 \\
    \tparam{Total parameters} & 983.6M \\
    \bottomrule
  \end{tabularx}
\end{table}

\suppinlinehead{Model architecture.}
Our model is built around a waveform-patch DiT generator.
The noisy target waveform is tokenized with non-overlapping 768-sample patches, so one generator token corresponds to 32~ms at 24~kHz.
The output layer predicts clean waveform values at the same patch resolution, and a non-overlapping unpatchify operation reconstructs the waveform in the sample domain.

\begin{table}[t]
  \centering
  \small
  \renewcommand{\arraystretch}{1.05}
  \newcommand{\tparam}[1]{\hspace{0.6em}#1}
  \caption{Optimization, loss, and sampling settings used for our experiments.}
  \label{tab:supp_train}
  \begin{tabularx}{\linewidth}{@{}p{0.48\linewidth}X@{}}
    \toprule
    Parameter & Value \\
    \midrule
    \multicolumn{2}{@{}l}{\textit{Optimization and batching}} \\
    \tparam{Hardware} & 4 NVIDIA A800 GPUs \\
    \tparam{Batch type} & Dynamic frame-wise batching \\
    \tparam{Optimizer} & Muon (DiT backbone), AdamW (others) \\
    \tparam{Learning rates} & $10^{-3}$ (Muon) / $5\times10^{-5}$ (AdamW) \\
    \tparam{AdamW betas} & $(0.9, 0.95)$ \\
    \tparam{Warmup} & 20k updates \\
    \tparam{Weight decay} & 0 \\
    \tparam{Gradient clipping} & 1.0 \\
    \midrule
    \multicolumn{2}{@{}l}{\textit{Flow matching and training losses}} \\
    \tparam{Noise scale} & 0.1 \\
    \tparam{Mask ratio} & $[0.7, 1.0]$ \\
    \tparam{Corruption schedule} & Logit-normal $(-0.4, 0.8)$ for eq. updates $< 240$k; uniform $[0, 1]$ for eq. updates $\ge 240$k \\
    \tparam{Prompt-audio drop prob.} & 0.3 \\
    \tparam{Prompt-audio + text drop prob.} & 0.2 \\
    \tparam{REPA teacher} & WavLM-base+ \\
    \tparam{REPA source layer} & 18th DiT block \\
    \tparam{REPA projection width} & 2048 \\
    \tparam{REPA loss weight} & $\lambda_1 = 0.0025$ \\
    \tparam{VAPA loss weight} & $\lambda_2 = 4 \times 10^{-4}$ \\
    \tparam{STFT FFT sizes} & 1024, 2048, 512 \\
    \tparam{STFT Hop sizes} & 128, 256, 64 \\
    \tparam{STFT Window sizes} & 512, 1024, 256 \\
    \midrule
    \multicolumn{2}{@{}l}{\textit{Sampling}} \\
    \tparam{Solver} & Heun \\
    \tparam{NFE} & 50 \\
    \tparam{CFG} & 3.5 \\
    \tparam{EMA} & 0.9999 \\
    \tparam{Guidance interval} & $[0,1]$ \\
    \tparam{Sway coefficient} & $-1.0$ \\
    \bottomrule
  \end{tabularx}
\end{table}

\suppinlinehead{Text conditioning and backbone.}
The text condition is represented at the character level.
Before interacting with audio, the character sequence is first refined in its own text stream by four ConvNeXt blocks with embedding dimension 512, expansion ratio 2, and depthwise kernel size 7.
This separate text modeling stage is used to organize the text condition before joint modeling with long audio sequences.
The resulting text tokens are then inserted as in-context tokens before the audio stream, yielding a single sequence in which text provides left-context guidance while the noisy waveform and the prompt-audio representations remain in the main generation stream.
The unified sequence is then processed by a 32-block DiT backbone with hidden size 1280, 16 attention heads, and MLP ratio 4.0.
Table~\ref{tab:supp_arch} summarizes the principal backbone hyperparameters.

\suppinlinehead{Optimization and batching.}
During training, we use Muon on the DiT parameters and AdamW on the remaining parameters\footnote{https://github.com/KellerJordan/Muon}.
The corresponding learning rates are $10^{-3}$ and $5\times10^{-5}$, with AdamW betas $(0.9, 0.95)$, zero weight decay, 20k warmup updates, and gradient clipping of 1.0.
For models without REPA, the batch size is 19200 waveform patches; when REPA is enabled, the budget is reduced to 9600 because of the additional frozen-teacher branch.
To keep comparisons fair, all update counts are reported as equivalent updates, obtained by normalizing runs with different batch budgets to the same seen-data scale.

\suppinlinehead{Training objective.}
The model directly predicts the clean waveform and is trained with a masked flow-matching objective.
For each training example, a random target span is selected as the region to generate, while the remaining waveform serves as the masked-speech condition together with the full text condition.
For each utterance, the masked region covers between 70\% and 100\% of the waveform, yielding a speech-infilling objective with long missing spans.
Classifier-free guidance is trained by dropping the prompt-audio condition with probability 0.3 and jointly dropping prompt audio and text with probability 0.2.
Table~\ref{tab:supp_train} summarizes the principal optimization, loss, and sampling hyperparameters.

\end{document}